\documentclass[twocolumn]{IEEEtran}
\usepackage{latexsym}
\usepackage[dvips]{graphicx}
\usepackage{amsmath}
\usepackage{amssymb}
\usepackage{subfigure}

\begin{document}

\title{Computational Complexity of Continuous Variable Quantum Key
Distribution\thanks{%
This work was supported in part by the National Fundamental Research Program
of China under Grant No 2006CB921900, National Natural Science Foundation of
China under Grants No. 60537020 and 60621064, Innovation Fund of the
University of Science and Technology of China under Grants No. KD2006005 and
the Knowledge Innovation Project of the Chinese Academy of Sciences (CAS)}}
\author{Yi-Bo Zhao, You-Zhen Gui, Jin-Jian Chen, Zheng-Fu Han and Guang-Can
Guo\thanks{%
All the authors are with the Key Lab of Quantum Information,
University of Science and Technology of China, (CAS), Hefei, Anhui
230026, China (e-mail: zfhan@ustc.edu.cn).}} \maketitle

\begin{abstract}
The continuous variable quantum key distribution has been considered to have
the potential to provide high secret key rate. However, in present
experimental demonstrations, the secret key can be distilled only under very
small loss rates. Here, by calculating explicitly the computational
complexity with the channel transmission, we show that under high loss rate
it is hard to distill the secret key in present continuous variable scheme
and one of its advantages, the potential of providing high secret key rate,
may therefore be limited.
\end{abstract}

\begin{keywords}
Computational complexity, Continuous variable, Error correction,
Quantum key distribution, Reconciliation.
\end{keywords}

\section{Introduction}

Due to its potential for achieving high modulation and detection speed,
continuous variable (CV) quantum key distribution (QKD) has recently
attracted more and more attention. Compared to single photon counting
schemes, CVQKD does not require single photon sources and detectors which
are technically challenging now. The CVQKD schemes typically use the
quadrature amplitude of light beams as information carrier, and homodyne
detection rather than photon counting. Some of these schemes use
non-classical states, such as squeezed states \cite{1} or entangled states
\cite{3}, while others use coherent states \cite{7,8,9,10}. Because the
squeezed states and entangled states are sensitive to losses in the quantum
channel, coherent states are much more attractive for long distance
transmission. To improve the performance of the CVQKD against the channel
loss, Grosshans et al proposed a reverse reconciliation (RR) protocol \cite%
{14}. In the traditional direct reconciliation protocol, Alice sends Bob the
quantum state and also sends the reconciliation information later\footnote{%
In the following we use the conventional appellation. Alice is the quantum
state sender and Bob is the quantum state receiver.}. Finally, Bob obtains
Alice's data without any error. However, in the reverse reconciliation
protocol, the quantum state is sent by Alice to Bob, but the reconciliation
information is sent by Bob to Alice. Finally, Alice gets Bob's received data
with no error.

Table-top experimental setups that encode information in the phase and
amplitude of coherent states has been demonstrated \cite{11,symul}, and
recent experiments have shown the feasibility of CVQKD in optical fibers up
to a distance of 55km \cite{12,13}, but without obtaining the final secret
keys.

Unlike the single photon QKD schemes, many CVQKD schemes utilize the
inertial quantum noise to protect information from Eve's attack \cite{11,15}%
. However, at the same time the quantum noise also causes errors between two
legitimate communicators, Alice and Bob. It is widely believed that through
the classical error correction the errors between Alice and Bob can be
corrected and partial information can still be kept in secret. Nevertheless,
in the QKD we should not only guarantee that the errors between Alice and
Bob can be corrected, but also ensure that Eve cannot know all the
information of the secret keys after the error correction. In the following
we will see that such requests actually pose a challenge to the error
correction. After showing that the lower bound of the block size of the
required error correction for the QKD is inversely proportional to the
square of the secret information carried by per element, we illustrate that
the reconciliation is a big hindrance to the CVQKD.

In principle, if the mutual information between Alice and Bob is larger than
that between Eve and Bob, Alice and Bob can establish unconditional secure
secret keys by the RRCVQKD protocol \cite{Secret sharing 1,Secret sharing
2,Public}. However, in practice we should also consider the feasibility.
Before establishing the secret keys, the errors contained in Alice and Bob's
variables should be corrected. Then certain amount of error correction
information\footnote{%
The error correction information is actually the check bit of the redundant
coding. In the QKD, the check bit can be separately sent.} should be
publicly exchanged \cite{15,20,21}. At the same time, while received this
public error correction information, Eve can get more information about
Alice and Bob. To guarantee the security we should ensure that Eve cannot
totally know Alice and Bob's key. Then the amount of publicly exchanged
information should be below certain threshold. Suppose before the error
correction the mutual information rate per key element between Alice and Bob
is $I_{AB}$ and that between Eve and Bob is $I_{EB}$, after the error
correction the mutual information rate per key element between Alice and Bob
is $I_{AB}^{\prime }$,\ and the amount of the published effective\footnote{%
There may be much auxiliary information exchanged, which however does not
bring any useful information about the key to Eve. Here the effective means
the information that may leak Eve useful knowledge about the key.} error
correction information rate per key element is $R$. Here we can see that
before the error correction the theoretical amount of secret key rate they
can generate is $\Delta I=I_{AB}-I_{BE}$-bit. After the error correction the
mutual information rate per key element between Alice and Bob becomes $%
I_{AB}^{\prime }\leq I_{AB}+R$ and that between Eve and Bob becomes $%
I_{BE}+R $.\ Then if Alice corrects her errors and finally shares Bob's bits
the secret key rate after the error correction becomes $\Delta
I-(R-I_{AB}^{\prime }+I_{AB})$. Here we can see that to generate pure secret
keys finally, it should be guaranteed that $R-(I_{AB}^{\prime
}-I_{AB})<\Delta I$. Set $I_{AB}^{\prime }-I_{AB}=R^{\prime }$. Therefore,
in practice we should use less than $R^{\prime }+\Delta I$-bit information
for per key element to correct errors that can only be corrected with more
than $R^{\prime }$-bit information for per key element. While the $\Delta
I/I_{AB}$\ exponentially decreases with the increase of the transmission
distance, the practical error correction should exponentially approach the
Shannon limit. In some literature, they defined a reconciliation efficiency $%
eff$ to evaluate the reconciliation \cite{11,14,15}. Here $1-eff$ actually
correspond to $(R-R^{\prime })/I_{AB}$. Also, in Ref. \cite{Matthias}, M.
Heid et al have evaluated the influence of the reconciliation efficiency to
the binary modulated CVQKD. From their result, it can be seen that if the
reconciliation efficiency is not 1, the maximum achievable distance of the
CVQKD will be reduced. In the following we will show the relationship
between the achievable distance and the computational complexity.\ \

\section{Difficulty with reverse reconciliation CVQKD: an example}

Here, we consider a specific example of Gaussian-modulated RRCVQKD \cite{11}%
. In the reverse reconciliation protocol, the quantum state is sent by Alice
to Bob but the reconciliation information is sent by Bob to Alice. Finally
the secret key rate is given by the difference between the mutual
information rate between Alice and Bob and that between Eve and Bob. After
the quantum communication, some CVQKD schemes directly provide the binary
keys \cite{7}, while the Gaussian-modulated CVQKD scheme only provide the
continuously distributed key elements \cite{11} that should be converted
into common binary bits. Here we only discuss the latter one. It has been
proposed that this conversion can be achieved by reconciliation.
Nevertheless, how to realize proper reconciliation is still an open problem.
One existing reconciliation protocol was suggested by G. Van Assche \textit{%
et al} \cite{15}, in which they subtly combined the quantization with the
sliced error correction. They quantize the continuous-variable into binary
keys at first and then do the error correction to those keys. Different from
single photon schemes, in the Gaussian-modulated RRCVQKD one, most of the
transmitted information can be known by Eve and only a very small portion
can be kept in secret. The signal to noise ratio (SNR) between Alice and Bob
is only slightly higher than that between Eve and Bob, and the difference of
the two may be overwhelmed by the fluctuation of noise. Then, the distilling
the secret key will amount to looking for a needle in a haystack. For
example, let us take the line loss to be 20dB (100km fiber with an
attenuation coefficient of 0.2dB/km), under which the maximum secret
information per key element carries is about 0.007-bit \cite{11} (i.e., on
average, 1-bit secret key requires 1/0.007=143 key elements). Suppose the
variance of Bob's measurement is $2N_{0}$, under which Alice's modulation
variance is $100N_{0}$, where $N_{0}$ denotes the variance of the vacuum
noise. Then the mutual information between Alice and Bob is 0.5-bit and that
between Eve and Bob is 0.493-bit. On estimating Bob's results, the
difference between Alice's and Eve's noise is about $0.01N_{0}$, which can
be calculated from Ref. \cite{11}. To distill the secret key, at first those
associated key elements should be converted into binary keys. The
quantization will certainly induce information loss, although the loss can
be exponentially reduced. To ensure the lost information to be less than
0.007-bit per key element, the reconciliation should be precise enough to
distinguish the slight noise change of $0.01N_{0} $, so each key element
should be converted into at least $0.5\log _{2}(2N_{0}/0.01N_{0})\approx 4$
independent binary digits. After the conversion, Alice and Bob at least get
4 binary digits from one key element and the secret information contained by
them will be certainly less than 0.007-bit. Since the information per
element carry is 0.5-bit, to correct bit-errors, at least 3.5-bit additional
information should be exchanged for one key element. Then, on average, 143$%
\times $4=572 binary digits contribute to less than one secret key and at
the same time 143$\times $3.5$\approx $500-bit additional information should
be exchanged. Therefore, the distillation is equivalent to extracting a
secret bit at least in 1072 binary digits, which is quite impractical as the
error fluctuations in 572 digits may easily overwhelm the 1-bit secret key
information to be distilled.

To distill the secret key, errors in quantized bits should be corrected. In
theory, the errors in 572 bits require at least 500-bit additional
information to correct, but the secret information carried by 572 bits is
less than 1-bit and Eve can use less than 501-bit additional information to
know all of Bob's information, which means that if 501-bit error correction
information is exposed there may be no secret information kept from Eve, so
to correct the errors in 572 bits and to ensure corrected bits still carry
secret information, the publicly exposed reconciliation information should
not exceed 501-bit. Then the practical error correction should use less than
501/500=1+0.2\% times the theoretical minimum information to correct the
errors at a high bit error rate (BER). While in Ref. \cite{20,21}, even at a
BER as small as 3\%, this ratio can be enhanced to 120\%, much higher than
1+0.2\%, but still pose a technical challenge. In fact, to improve the
practical channel information capacity is a hot topic in classical
communications \cite{18}. One major goal is to correct the errors with as
little redundant information as possible. Up to now, even under ideal
situations (e.g., in the standard channel, at fixed coding rate, under the
optimized SNR and distinguishing the BER of each bit), present
state-of-the-art error correction schemes, such as Low-Density Parity-Check
(LDPC) \cite{18} and Turbo \cite{19} codes, may in theory only get
information with low BER from 1.001 times (0.0045dB) of the theoretical
minimum resource \cite{18}. While in practice, this theoretical performance
is still hard to reach at present. In the above example, Alice should get
information from less than 572+501=1073 binary digits without any error,
while in theory this information should be obtained from at least 1072
binary digits. The ratio of them is 1073/1072=1.0009 or 0.0040dB, smaller
than 0.0045dB, which is a huge challenge to the error correction technique.
In practice, the required performance of error correction will be much
higher. To our knowledge there is no such error correction that can meet its
requirement.

In Refs. \cite{11,15}, they sorted those uncorrelated bits approximately
with the same BER into one group and correct errors respectively. It is
possible that BERs of some groups are smaller than the average one, so their
errors are more easily to correct than the above analysis. However, there
are certainly some groups the BER of which are higher than the average one
and require much more complex error correction. Then on average its required
error correction may be more difficult to realize than the above analysis.
Actually, in Ref. \cite{11} they directly exposed these bits with a high
BER, whereas whether such operation affects the security requires further
discussion.

In the above, we have discussed a limit case, under which the quantization
can be ideal and the correlation as well as the asymmetry can be neglected.
In fact, such limit case cannot be reached and each of the converted digit
may be correlated with each other and the channel may be binary asymmetric,
so one continuous-element should be converted into many more digits. Also,
the correlation and the asymmetry will greatly increase the complexity and
reduce the efficiency of the error correction. Therefore, in practice,
things will be much more difficult than what we have shown in the above
example.

\section{Computational complexity of the required error correction}

\subsection{Simple model}

In the following, we will calculate explicitly the computational complexity
of the error correction including secret key information carried by single
digit. Here we will start from a binary symmetric model, and then extend it
to more general case. Later we will show that such model can be applied to
current CVQKD scheme. Suppose after many communications and processes, Alice
and Bob get binary strings $M_{A}=(M_{A1},\ldots ,M_{Am})^{T}$ and $%
M_{B}=(M_{B1},\ldots ,M_{Bm})^{T}$ of length $m$, respectively. In the
following, it is assumed that the channel is binary symmetric, i.e., the
binary strings contain equal probabilities of 0's and 1's, the bits in the
strings are uncorrelated. Also, it is assumed that the error correction
information is sent by Bob to Alice and with this error correction
information Alice can correct all of the errors only when the number of
errors is below a threshold value \cite{16}. We define $e_{AB}=\Pr
(M_{Ai}\neq M_{Bi})$ as the BER and $S(M_{B}|E)$ as the entropy rate per bit
of Bob's string conditioned on Eve's state. Then the maximum amount of
secret key that can be distilled by Alice and Bob is $%
m[S(M_{B}|E)-H(e_{AB})] $-bit, where $H(x)=-x\log _{2}x-(1-x)\log _{2}(1-x)$
is the Shannon entropy. To share the common secret key, Alice and Bob should
eliminate the errors in their final binary string. Then, they can use, for
example, error correction, to tolerate certain amount of errors. Suppose
Alice can correct at most $me_{rec}$ errors with Bob's error correction
information, where $e_{rec}$ is an introduced parameter. Then $mH(e_{rec})$%
-bit additional information is required and after the error
correction, the maximum amount of secret keys becomes
$m[S(M_{B}|E)-H(e_{rec})]$-bit. While the average amount of errors
between Alice and Bob is $me_{AB}$, the actual amount of errors will
fluctuate, and the fluctuation obey the binomial distribution with
variance $me_{AB}(1-e_{AB})$ \footnote{Under the symmetric binary
channel condition, the distribution
of the number of errors obey the binominal distribution: $%
P(n)=C_{m}^{n}e_{AB}^{n}(1-e_{AB})^{m-n}$, where $n$ describes the
number of errors.}. Here, we introduce parameter $\lambda $ and
suppose the number of practical errors is $m\lambda $. Then the
binomial distribution of errors between Alice and Bob can be
approximated by a Gaussian:%
\begin{eqnarray}
P(\lambda ) &\approx &\frac{m}{\sqrt{2\pi me_{AB}(1-e_{AB})}}  \notag \\
&&\times \exp [-\frac{(me_{AB}-m\lambda )^{2}}{2me_{AB}(1-e_{AB})}]
\label{e1}
\end{eqnarray}

Only when $\lambda \leq e_{rec}$ can the errors be corrected, so the
probability that Alice correct her errors is%
\begin{eqnarray}
\beta &=&\int_{0}^{e_{rec}}P(\lambda )d\lambda \approx \int_{-\infty
}^{e_{rec}}\frac{m}{\sqrt{2\pi me_{AB}(1-e_{AB})}}  \notag \\
&&\times \exp [-\frac{(me_{AB}-m\lambda )^{2}}{2me_{AB}(1-e_{AB})}]d\lambda
\label{e2}
\end{eqnarray}

To share common secret key with high probability, Alice and Bob should
guarantee that $\beta \rightarrow 1$. To distill the secret key, they should
maintain $S(M_{B}|E)>H(e_{rec})$. Then it requires $m$ to be sufficiently
large. For given probability $\beta $, error correction criteria $e_{rec}$
and BER $e_{AB}$, the length $m$ can be given by:%
\begin{equation}
m=\frac{e_{AB}(1-e_{AB})}{(e_{rec}-e_{AB})^{2}}Q^{2}(1-\beta )  \label{e3}
\end{equation}%
where $Q(x)$ is defined as the solution of the equation $1/\sqrt{2\pi }%
\int_{Q(x)}^{\infty }\exp (-y^{2}/2)dy=x$ and we have supposed $%
e_{rec}>e_{AB}$.

Here, we introduce a parameter $e_{c}$ and suppose $H(e_{c})=S(M_{B}|E)$.
Then from the requirement $S(M_{B}|E)>H(e_{rec})$, we know that $%
e_{c}>e_{rec}$.\ In practice, $e_{c}$ and $e_{AB}$ are determined by the
quantum channel, the $\beta $ is determined by practical requirements, so
Alice and Bob need to choose proper $e_{rec}$ and $m$ to satisfy the Eq. (%
\ref{e3}). Set $\delta I=H(e_{c})-H(e_{AB})$, where $\delta I$ denotes the
secret rate per digit. Then we have%
\begin{eqnarray}
m &>&e_{AB}(1-e_{AB})(\log _{2}\frac{1-e_{AB}}{e_{AB}})^{2}(\frac{1}{\delta I%
})^{2}  \notag \\
&&\times Q^{2}(1-\beta )  \label{e8}
\end{eqnarray}%
where we have used the result that $e_{rec}<e_{c}$ and $e_{c}-e_{AB}\approx
\delta I/\log _{2}\frac{1-e_{AB}}{e_{AB}}$.

The value of $m$ will determine the computational complexity of the error
correction. For the common error correction code, the computational
complexity is of order $O(m^{2})$ \cite{18}.

In the above we have discussed the binary symmetric case. In the practical
case the channel may be asymmetric and the BER of different bit may be
different. Under such cases whether the error can be corrected is not only
determined by the number of errors, but also determined by the BER of each
bit. There is still a threshold for such case, which is determined by the
amount of the published error correction information, but now whether
certain amount of errors can be corrected or not becomes probabilistic \cite%
{18,19,Gilbert}. If the number of errors is larger than this threshold, all
the errors can be corrected with high probability. On the contrary, if the
number of errors is smaller than this threshold, the errors can be corrected
with comparatively low probability. The larger the number of the errors, the
smaller the probability that all errors can be corrected is. To guarantee
that possible errors can be corrected with much high probability, Alice and
Bob should also ensure that the actual number of the errors is smaller than
this threshold with a high probability. The upper bound of this threshold is
determined by Alice's conditional entropy conditioned on Bob and the secret
key rate. Then the problem becomes similar to that under the binary
symmetric case. According to the large number theory, the distribution of
the number of errors can also be approximated by the Gaussian distribution
under this case. Then by the same way used for the binary symmetric case, we
can also prove that under the case that the channel is asymmetric or the BER
of different bit is different, the minimum block size of the required error
correction code satisfies
\begin{equation}
m\propto (\frac{1}{\delta I})^{2}.  \label{e9}
\end{equation}%
Here we have not written Eq. (\ref{e9}) explicitly into terms of the BER and
$Q^{2}(1-\beta )$, because under the asymmetric case it is difficult to give
a general form to the variance of the Gaussian approximation and actually
only the relationship between $m$ and $\delta I$ is essential for us. In the
following we will discuss the relationship between the minimum block size
and the channel transmission.

In the reconciliation given in the Ref. \cite{15}, they grouped the bits
into several strings, so that in one string different bit corresponds to
different key element. Then if Eve does the individual attack\footnote{%
Generally, Eve's attack can be separated into three classes, individual
attack, collective attack and coherent attack. The individual attack means
that Eve attacks each signal system separately with the same method and
later measures her quantum state right after the sifting step. The
collective attack means that Eve attacks each signal system separately with
the same method but can do arbitrary measurement after all of the steps,
including the reconciliation, privacy amplification and the encryption. The
Coherent attack is the most general attack, where Eve can attack all of the
signals together and do the measurement at the end of the protocol.} or the
collective attack, the bits in one string can be uncorrelated. Also, since
Eve attacks each signal by the same method, the distribution of each key
element is identical, so that the Gaussian approximation can be used to
describe the distribution of the number of errors in one string. Therefore
under the individual attack or the collective attack case the above model
can be applied to the reconciliation given in Ref. \cite{15}. There may be
other reconciliation method, in which the bits in one string are totally
correlated with each other, so that the above model can not be applied.
However, the efficiency of current error correction code will be much low if
the bits in one string are correlated. Actually, under the current
technology, the computational complexity of the error correction will be
largely increased if we make the bit correlated. Therefore in the following
we only discuss the uncorrelated case.

In the above we have discussed the relationship between the minimum block
size and the secret key rate. For the binary symmetric case, we gave the
explicit expression for the minimum block size in Eqs. (\ref{e3}) and (\ref%
{e8}). Since the binary symmetric case is the easiest condition to handle,
using this result we can give a rough estimation to the minimum block size.
We also considered the non binary symmetric case. However, we have not given
the explicit expression for the block size under this case. Because under
such case the minimum block size depend on the concrete properties of the
channel, it is difficult for us to give a general form. At least using Eq. (%
\ref{e9}) we can give a rough estimation for the order of the computational
complexity if some experimental results are given.

\subsection{Computational complexity of present schemes}

In single photon schemes, after the quantum communication, almost all the
information of the distributed binary string is kept in secret from Eve,
i.e., $\delta I\rightarrow 1$, so the required $m$ can be very small.
However, in CVQKD, most of the information may be tapped by Eve, and
consequently, the $\delta I$ is very close to $0$, so that the required $m$
will be very large. We can give a rough estimation to the example given in
Sec. II. We suppose the channel is binary symmetric. At first we roughly
estimate the BER\ and the secret key rate per binary digit of a typical
string. Alice's conditional entropy rate conditioned on Bob's variable is
3.5-bit per key element. Since Alice and Bob map their key element into 4
digit, we estimate BER $e_{AB}$ of a typical converted binary string by $%
4H(e_{AB})=3.5$. Thus $e_{AB}\approx 0.29$. Also we suppose the secret
information each digit carries is 0.007/4=0.00175-bit\footnote{%
The secret key rate 0.007-bit per key element is obtained under the
assumption that Eve is classical.}. If we set $1-\beta =10^{-7}$, then $%
Q^{2}(1-\beta )\approx 27$. Using Eq. (\ref{e3}), we obtain $m>10^{7}$. That
means, to share common secret key with high probability and to restrain the
probability that Eve gets their key, Alice and Bob should set the criteria
of the error correction at least to $10^{7}$ digits. Then, even with ideal
error correction technique, the required block size should be larger than $%
10^{7}$. For present LDPC coding, only under the simulation condition and
with a very simple code, can this block size be realized \cite{18}. Because
we suppose that the channel is binary symmetric and many other factors have
not been taken into account, the estimated result may be far away from the
practical one. In the following we will give an estimation based on current
experiment.

From the Ref. \cite{11} we know that for the RRCVQKD if there is no excess
noise and Eve is classical the secret key rate per key element is
proportional to $1/\eta $, where $\eta $ denotes the channel transmission.
Then we can safely assume that after the conversion, the secret key
information carried by single binary bit is proportional to $1/\eta $. Since
the computational complexity generally is proportional to the square of the
block size \cite{18},\ from Eq. (\ref{e9}) we conclude that the minimum
computational complexity of the error correction is at least of order $%
O(m^{2})\approx O(1/\delta I^{4})\approx O(1/\eta ^{4})\approx
O[\exp (0.18L)]$, where $L$ denotes the transmission distance in
units of km, and the fiber attenuation efficiency has been chosen to
be 0.2dB/km. The authors of Ref. \cite{11}, can distill the secret
key within 3.1dB loss ($\eta \approx 0.5$, 15km fiber), even though
high performance error correction was used and Eve was assumed to be
classical. While under the condition of 10dB loss ($\eta =0.1$, 50km
fiber), the required error correction will be at least 600 times
more complex. Under the 100km fiber case, it will be at
least $6\symbol{94}10^{6}$ times more complex\footnote{%
This relationship is of the minimum block size. It is possible that the
efficiency of the reconciliation used in Ref. \cite{11} is low and after
some improvement the practical computational complexity under 3.1dB loss can
be largely reduced. Therefore, this relationship does not accurately
represent the practical one.}. Then, whether the secret key can be distilled
becomes doubtful. Moreover, as the error correction becomes very complex,
the practical computation speed of the algorithm will certainly become very
low. Therefore, even though the secret key can be distilled and the physical
modulation and detection rate can be very high, the practical secret key
rate, which will be limited by computational complexity, cannot be actually
improved.

\section{Remarks}

The purpose of this letter is to show a hindrance of the secret key
distillation for the CVQKD under the high loss condition, so in many
sections we just discussed an ideal case. In the example of Sec. II we
supposed an ideal quantization that can convert continuous-variables into
symmetric and independent bit string which is the easiest to handle.
Consequently, if we show that even under this ideal condition, the error
correction is prohibitive, then it will be much more prohibitive under the
practical case. Moreover, here we not only discussed the error correction,
but also considered the security. Although it is widely believed
(Gilbert-Varshanov conjecture) that general decoder can only correct half
the errors which can be corrected by an ideal error correction code \cite%
{Gilbert}, in the security analysis we should assume that Eve has the most
powerful error correction and her ability should only be restricted by the
theoretical limit rather than the wide belief. We can see that even under
the condition that Alice can correct all the errors an ideal error
correction code can correct, the computational complexity is still very
exaggerated. If the practical error correction is the widely believed one,
the practical reconciliation will be much more prohibitive.

It also should be noted that we supposed Eve's attack is individual attack
or collective attack, so by the reconciliation method given in Ref. \cite{15}
we can ensure that the bits in Alice and Bob's strings are uncorrelated. If
we take the most powerful attack, coherent attack, into account, the
computational complexity will be much more complex$\footnote{%
The individual attack and the collective attack are the subclass of the
coherent attack.}$. Here we were going to show the impossibility of the
secret key distillation at long distance transmission, so we only considered
the individual and collective attack case. If the secret key cannot be
distilled under this limit condition, it certainly cannot be distilled under
the coherent attack condition.

In the examples of Sec II and III we were discussing the pure RRCVQKD that
is without any other tactic, such as post-selection, assisted. For this pure
RRCVQKD, it has been shown that the mutual information between Alice and Bob
is certainly larger than that between Eve and Bob and thus it is expected to
provide high secret key rate. Here we showed that if other tactics are not
employed, the minimum computational complexity is certainly exponentially
proportional to the transmission distance.

\section{Conclusion}

In conclusion, we have discussed in detail the computational complexity of
the error correction required by the CVQKD. It has been shown that if other
tactics, such as post-selection \cite{9}, is not introduced, the minimum
computational complexity of the RRCVQKD is exponentially proportional to the
transmission length. For single photon schemes, the single photon source and
detector are the main limitations to the transmission distance and the
secret key rate \cite{22}, while for present CVQKD we have shown that the
computational complexity may be the key hurdle. Although the CVQKD solve the
problem of the light source and detection, it brings a new problem,
computational complexity. With the progress of the computer and new error
correction technique, the computational complexity of the RRCVQKD may no
longer be a problem, but under existing techniques, its strict requirements
are difficult to satisfy.

\begin{center}
\textbf{Acknowledgments}
\end{center}

Special thanks are given to Ling-An Wu, H. Pu and P. Grangier. We also thank
F. Zhang and J. Ma for the discussions on the error correction.


\end{document}